\documentclass[acmtog, authorversion]{acmart}
\acmSubmissionID{715}

\usepackage{booktabs}
\usepackage{graphicx}
\usepackage{enumitem}
\usepackage{soul}
\usepackage{dsfont}
\usepackage{gensymb}
\usepackage{microtype}
\usepackage{pifont}
\usepackage{threeparttable}
\usepackage{collect}
\usepackage{xspace}
\usepackage[capitalise]{cleveref}
\usepackage{natbib}
\usepackage[normalem]{ulem}
\usepackage{xfrac}
\usepackage{multirow}

\definecolor{darksalmon}{rgb}{0.91, 0.59, 0.48}

\definecolor{salomon}{rgb}{1, 0.6, 0.5}
\definecolor{pastelblue}{RGB}{0,153,204}
\newcommand{\NEW}[1]{\textcolor{black}{#1}}

\newcommand{\eg}{e.g.,\ }
\newcommand{\ie}{i.e.,\ }

\newcommand{\refFig}[1]{Fig.~\ref{fig:#1}}

\newcommand{\refTab}[1]{Tab.~\ref{tab:#1}}

\newcommand{\refSec}[1]{Sec.~\ref{sec:#1}}

\newcommand{\refEq}[1]{Eq.~\ref{eq:#1}}

\newcommand{\dataset}[1]{\textsc{#1}}

\definecollection{mymaths}

\newcommand{\mymath}[2]{
    \newcommand{#1}{\TextOrMath{$#2$\xspace}{#2}}
    \begin{collect}{mymaths}{}{}{}{}
    #1
    \end{collect}
}

\mymath{\scalespace}{\bar{I}}
\mymath{\outcolor}{\mathbf{c}}
\mymath{\coord}{\mathbf{x}}
\mymath{\bandwidth}{\omega}
\mymath{\minfrequency}{{\bandwidth_\textrm{min}}}
\mymath{\maxfrequency}{{\bandwidth_\textrm{max}}}
\mymath{\scale}{s}
\mymath{\maxscale}{{\scale_\textrm{max}}}
\mymath{\discretecoord}{{\mathbf{x}_i}}
\mymath{\resolution}{N}
\mymath{\maxresolution}{{N_\textrm{max}}}
\mymath{\patchresolution}{{\resolution_p}}
\mymath{\patchcoord}{{\coord_p}}
\mymath{\patchscale}{{\scale_p}}
\mymath{\generator}{G}
\mymath{\discriminator}{D}
\mymath{\latentcode}{\mathbf{z}}
\mymath{\generatedscalespace}{{\scalespace_\generator}}
\mymath{\mappingnetwork}{M}
\mymath{\synthesisnetwork}{S}
\mymath{\fourierfeature}{\mathbf{f}}
\mymath{\fourierfeaturedim}{{d_\fourierfeature}}
\mymath{\fourierfeaturefreq}{\boldsymbol{\omega}}
\mymath{\transform}{g}
\mymath{\freqset}{F}
\mymath{\setcount}{N}
\mymath{\basescale}{{\scale_\textrm{base}}}
\mymath{\weightfct}{w}
\mymath{\trainingiteration}{i}
\mymath{\resampling}{R}
\mymath{\patchdistance}{d}

\setcopyright{rightsretained}
\acmJournal{TOG}
\acmYear{2024} \acmVolume{43} \acmNumber{4} \acmArticle{131} \acmMonth{7}\acmDOI{10.1145/3658190}

\begin{document}

\title{Learning Images Across Scales Using Adversarial Training}

\author{Krzysztof Wolski}
\orcid{0000-0003-2290-0299}
\affiliation{%
 \institution{Max-Planck-Institut für Informatik}
 \city{Saarbrücken}
 \country{Germany}}
\email{kwolski@mpi-inf.mpg.de}

\author{Adarsh Djeacoumar}
\orcid{0009-0008-1919-450X}
\affiliation{%
 \institution{Max-Planck-Institut für Informatik}
 \city{Saarbrücken}
 \country{Germany}}
\email{adjeacou@mpi-inf.mpg.de}

\author{Alireza Javanmardi}
\orcid{0009-0008-4926-1566}
\affiliation{%
 \institution{Max-Planck-Institut für Informatik}
 \city{Saarbrücken}
 \country{Germany}}
\email{alireza.javanmardi@dfki.de}

\author{Hans-Peter Seidel}
\orcid{0000-0002-1343-8613}
\affiliation{%
 \institution{Max-Planck-Institut für Informatik}
 \city{Saarbrücken}
 \country{Germany}}
\email{hpseidel@mpi-sb.mpg.de}

\author{Christian Theobalt}
\orcid{0000-0001-6104-6625}
\affiliation{%
 \institution{Max-Planck-Institut für Informatik}
 \city{Saarbrücken}
 \country{Germany}}
\email{theobalt@mpi-inf.mpg.de}

\author{Guillaume Cordonnier}
\orcid{0000-0003-0124-0180}
\affiliation{%
 \institution{Inria, Université Côte d'Azur}
 \city{Sophia-Antipolis}
 \country{France}}
\email{guillaume.cordonnier@inria.fr}

\author{Karol Myszkowski}
\orcid{0000-0002-8505-4141}
\affiliation{%
 \institution{Max-Planck-Institut für Informatik}
 \city{Saarbrücken}
 \country{Germany}}
\email{karol@mpi-inf.mpg.de}

\author{George Drettakis}
\orcid{0000-0002-9254-4819}
\affiliation{%
 \institution{Inria, Université Côte d'Azur}
 \city{Sophia-Antipolis}
 \country{France}}
\email{george.drettakis@inria.fr}

\author{Xingang Pan}
\orcid{000-0002-5825-9467}
\affiliation{%
 \institution{Nanyang Technological University}
 \city{Singapore}
 \country{Singapore}}
\email{xingang.pan@ntu.edu.sg}

\author{Thomas Leimkühler}
\orcid{0009-0006-7784-7957}
\affiliation{%
 \institution{Max-Planck-Institut für Informatik}
 \city{Saarbrücken}
 \country{Germany}}
\email{thomas.leimkuehler@mpi-inf.mpg.de}

\begin{abstract}

\NEW{
The real world exhibits rich structure and detail across many scales of observation.
It is difficult, however, to capture and represent a broad spectrum of scales using ordinary images.
We devise a novel paradigm for learning a representation that captures an orders-of-magnitude variety of scales from an unstructured collection of ordinary images.
We treat this collection as a distribution of scale-space slices to be learned using adversarial training, and additionally enforce coherency across slices.
Our approach relies on a multiscale generator with carefully injected procedural frequency content, which allows to interactively explore the emerging continuous scale space.
Training across vastly different scales poses challenges regarding stability, which we tackle using a supervision scheme that involves careful sampling of scales.
We show that our generator can be used as a multiscale generative model, and for reconstructions of scale spaces from unstructured patches.
Significantly outperforming the state of the art, we demonstrate zoom-in factors of up to 256x at high quality and scale consistency.
}

\end{abstract}

\begin{teaserfigure}
	\includegraphics[width=\textwidth]{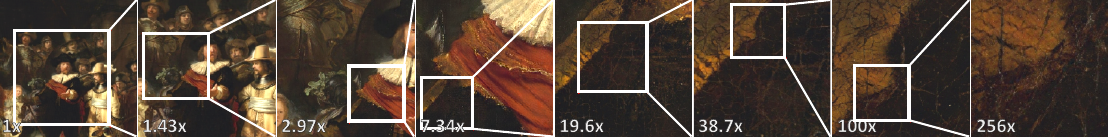}
    \vspace{-6mm}
	\caption{
    \NEW{
    Given an unregistered collection of image patches depicting an environment at vastly different scales, our approach uses adversarial training to obtain continuous and coherent scale spaces.
    Here, we showcase the reconstructed scale space of a painting, captured in its entirety, from the overall structure (1x) to the cracks in the oil paint (256x).
    Users can freely explore the scale space at interactive rates.
    }
    }
	\label{fig:teaser}
	\Description[TeaserFigure]{TeaserFigure}
\end{teaserfigure}



\maketitle

\section{Introduction}

The physical world exhibits a vast variety of scales, ranging from subatomic particles to galaxy clusters \cite{eames1968}.
A significant subset of these scales is within the reach of the human observer.
A rich visual representation of the world, therefore, needs to account for as many scales as possible \cite{lindeberg2013scale}, while
ordinary images can only capture a small slice of the scale spectrum due to two fundamental limitations:
They have \emph{finite extent} and \emph{finite resolution} \cite{koenderink1984structure}.

\begin{figure}
	\includegraphics[width=\linewidth]{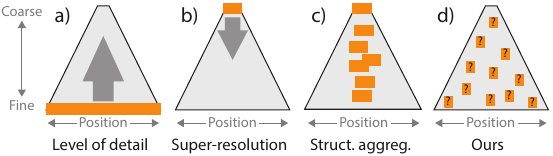}
	\caption{
	Different paradigms to obtain a multiscale image representation. Orange blocks indicate the location of the input data in scale space (trapezoid).
    \emph{a)} Level-of-detail methods require a full image at the finest scale and construct the scale space using low-pass filtering.
    \emph{b)} Super-resolution infers slightly finer scales from a coarse-scale image.
    \emph{c)} Approaches relying on structured aggregation assume registered images. 
    \emph{d)} Our approach relies on an \emph{unstructured} collection of low-resolution input images: The locations of the images in scale space are unknown (question marks in the orange blocks) and do not even necessarily depict the same scene. We nevertheless produce full coherent scale spaces.
	}
	\label{fig:positioning_scheme}
\end{figure}

Solutions to overcome these limitations can roughly be divided into three categories (Fig.~\ref{fig:positioning_scheme}a-c).
\emph{Level-of-detail} methods \cite{witkin1987scale,mallat1989theory} construct a set of coarser-scale versions from a given full high-resolution image (Fig.~\ref{fig:positioning_scheme}a), but obtaining the entire original image becomes infeasible for large scale spans.
In contrast, \emph{super-resolution} 
infers finer scales from a coarse-scale image (Fig.~\ref{fig:positioning_scheme}b), hallucinating plausible higher-frequency content \cite{wang2020deep,moser2023}, but seems to reach an upper limit of upsampling in the order of 10x.
Finally, methods that perform \emph{structured aggregation}~\cite{halladjian2019scale,xiangli2022bungeenerf} combine multiple images into a multiscale representation (Fig.~\ref{fig:positioning_scheme}c), but require dense capture and careful registration of images across all scales.

\begin{figure}
	\includegraphics[width=0.9\linewidth]{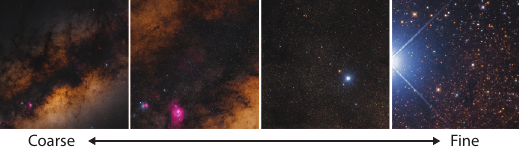}
	\caption{
	Typical samples from a multiscale dataset. The images have a fairly low resolution ($256 \times 256$ for us) and are unstructured, \ie we do not have information about relative 2D location, allowing uncomplicated capture or collection without the need for registration. \NEW{Images courtesy of Bartosz Wojczyński \shortcite{wojczynski2021}.}
	}
	\label{fig:dataset_examples}
\end{figure}

We introduce a novel approach for the construction of a multiscale image representation from a set of \emph{low-resolution, unstructured} images of a 2D environment (Fig.~\ref{fig:positioning_scheme}d). In particular, we consider images that have the same resolution but observe the environment at different scales, \ie finer-scale observations cover smaller patches (Fig.~\ref{fig:dataset_examples}). To alleviate the need for costly registration, \emph{we do not require information on the location of the patches}, but only an approximate indication of scale. Such data corresponds, for example, to remote sensing applications whose objective is to capture a geographical point of interest at scales as various as the different acquisition tools \eg satellites, airplanes, or drones flying at different altitudes. 
Note that the previous work discussed above cannot process this data.
The absence of a single high-resolution image does not allow level-of-detail, the scale spans we consider are orders of magnitude above the capabilities of super-resolution, and missing positional cues prevent structured aggregation.

\NEW{
We observe that an unstructured collection of image patches across scales constitutes a data distribution that can be learned using adversarial training~\cite{goodfellow2014generative}.
Our key contribution is a neural architecture and training paradigm that treats multiscale image patches as slices of an underlying continuous scale space and enforces coherency across space and scale.
This leads to two complementary training goals:
\emph{(i)} 2D slices of the generated scale-space(s) should match the distribution of the training data, and
\emph{(ii)} the generated scale space(s) should be coherent across all dimensions.
}

We build our representation upon an alias-free StyleGAN generator network~\cite{karras2021alias} augmented with a set of progressive Fourier features distributed across multiple layers that generate tailored latent frequency content across the scale spectrum.
Our training process is stable despite the wide range of scales and is specifically designed to enforce cross-scale consistency.

Once trained, we can interactively query our network with a position and scale level to obtain a corresponding generated scale-space slice. 
As shown in the companion video, it is possible to continuously zoom into \emph{any} location of the sample or pan over the image at a chosen scale.
Our generator synthesizes a single fixed-resolution image at a time but multiple adjacent samples can be seamlessly stitched together to yield coherent composites of up to several gigapixels at arbitrary scales.
We demonstrate magnifications of up to 256x, \ie one pixel in a 256x256 image can be enlarged to yield a full-resolution image with plausible high-frequency structure and details while being coherent across the continuous scale spectrum (\refFig{teaser}).

We demonstrate two applications of our method:
First, we show that we are able to aggregate the highly unstructured input into a coherent scale space,
\NEW{
\ie our approach produces a \emph{pseudo-reconstruction} of the underlying scale space by learning a regularized distribution of the input patches.
In doing so, our approach does not only implicitly register the patches, but also creates a compact representation, requiring 885x less parameters than the number of pixel values than an equivalent gigapixel images.
}
Second, we show that we can learn a \emph{generative model of scale spaces}, \ie provided with image patches from different environments, our model allows to draw multiple, independent yet consistent scale-space samples.

We evaluate our method on several satellite datasets, a multiscale dataset consisting of an unstructured collection of images from the internet, as well as synthetic datasets created by extracting multiscale patches from gigapixel images.
The latter provides us with ground-truth data for quantitative analysis.
Our datasets, code, and trained models can be found at \textcolor{pastelblue}{\url{https://scalespacegan.mpi-inf.mpg.de}}.

In summary, our contributions are 1) a novel paradigm based on adversarial training to obtain a compact continuous multiscale image representation from unstructured ordinary images, 2) a generator architecture and training methodology for stable and scale-consistent scale space generation, 3) the application of our approach for multiscale unstructured image aggregation and as a multiscale generative 
model, 4) an interactive rendering application (approx. 20\,fps) demonstrating the inference speed and data compression performance of our method.

\section{Related Work}

\subsection{Multiscale Representations}
\label{sec:related_representation}

The representation of signals at multiple scales has a long history in mathematics, signal processing, as well as computer graphics and computer vision.
Arguably the most concise description of phenomena at multiple scales is delivered by fractals \cite{mandelbrot1982fractal}. 
It provides a framework for modeling self-similarities and complexity across the (possibly infinite) scale spectrum, but is typically readily applicable only to a narrow class of signals.
Linear scale space theory \cite{iijima1959basic,witkin1987scale,lindeberg2013scale} aims at representing images at multiple scales by embedding them into a one-parameter family of progressively smoothed versions. 
Scale-space methods are omnipresent in computer vision from the earliest methods \cite{marr1980theory} till today \cite{lindeberg2022scale}.
Pyramids \cite{burt1981fast, williams1983pyramidal} are more compact multiscale representations, which perform scale discretization and spatial subsampling in addition to smoothing.
Wavelets \cite{daubechies1988orthonormal,mallat1989theory} represent signals across scales by constructing a multiscale basis.

Within the torrent of deep learning in the last decade, many neural continuous multiscale representations have been developed.
A broad range of visual-computing primitives has been considered, including images \cite{chen2021learning,xu2021ultrasr,paz2022multiresolution,Belhe2023Discontinuity}, geometry \cite{takikawa2021neural}, materials \cite{kuznetsov2021neumip}, radiance fields \cite{barron2021mipnerf,xiangli2022bungeenerf}, as well as general-purpose neural architectures \cite{lindell2022bacon,saragadam2022miner,shekarforoush2022residual,fathony2020multiplicative}.

All of the works listed above are designed to represent an original signal alongside its coarser-scale equivalents.
Consequently, they require explicit \emph{access to the entire signal at the finest scale}.
This poses a severe problem when considering an orders-of-magnitude variety of scales, as, in this case, the finest scale contains details that are hard to synthesize or capture.
In contrast, in this work, we develop a multiscale generative image model, which only requires an \emph{unstructured} collection of fairly \emph{low-resolution} images that capture \emph{patches} of an environment at different scales.
Similar to previous works \cite{barron2021mipnerf}, we employ progressive Fourier embeddings to steer our generator.

In the context of multiscale data visualization, exploration systems commonly blend between different representations best suited to convey information at a specific scale \cite{halladjian2019scale,klashed2010uniview,mohammed2017abstractocyte,tao2019kyrix}.
Similar ideas have been utilized in stitching-based variable-resolution image creation \cite{licorish2021adaptive}.
These approaches require registered images to be able to combine different sources of information, while our method relies on unstructured image collections.
We, too, support interactive exploration of the scale space of our samples, due to a lightweight generator design.


\subsection{Super-resolution}
\label{sec:related_superres}

With origins in image restauration and deconvolution \cite{wiener1949extrapolation,parker2010algorithms}, single-image super-resolution methods aim at increasing the resolution of an image while synthesizing plausible higher-frequency content.
Over the last years, feed-forward CNN-based approaches have established strong baselines for fixed-scale upsampling \cite{wang2020deep,chen2021pre,liang2021swinir,lim2017enhanced,shi2016real,zhang2018image,zhang2018residual,yang2020learning,lu2021masa}.
Arbitrary-scale super-resolution methods take the upsampling factor as an additional input \cite{hu2019meta,son2021srwarp,wang2021learning,wei2023super,song2023ope,Vasconcelos_2023_CVPR}, allowing them to synthesize a range of scales.
Generative models have been used as strong priors for super-resolution \cite{moser2023}, with a particular focus on GANs 
\cite{wang2018esrgan,wang2021real,Chan_2021_CVPR,menon2020pulse},
and, recently, diffusion models
\cite{wang2023unlimited,gao2023implicit,kawar2022denoising,wang2023exploiting,lin2023diffbir}.
Typical upsampling factors for super-resolution methods are in the order of 10x -- more than an order of magnitude less than what our approach can handle.


\begin{figure*}[]
	\includegraphics[width=\linewidth]{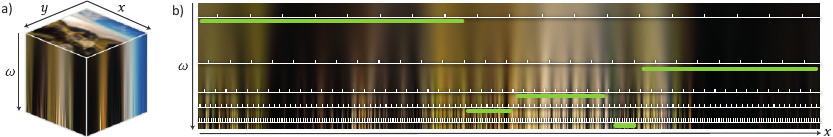}
	\caption{
    (\emph{a}) A scale space is a multiscale representation of an image. 
    It is a continuous function of spatial coordinates $\coord = (x, y)^T$ and bandwidth \bandwidth.
    Increasing \bandwidth introduces higher and higher frequencies.
    (\emph{b}) An $x$-\bandwidth-slice through the volume in \emph{a}).
    The resolution of spatial discretizations (white grids) needs to be adapted to a given \bandwidth to capture all frequency content.
    Input to our method is an unstructured collection of 2D image patches that sample the scale space (green bars).
    Each patch has a continuous location and scale.
    All patches have the same resolution ($\patchresolution=8$ in this visualization), which leads to different coverage of the spatial domain depending on their scale.
    Our method generates orders-of-magnitude scale spaces from this unstructured information.
    Notice that it is difficult to depict the actual resolution levels in a figure this size.
    We consider scale spaces, where the image at the top already requires a resolution of $256 \times 256$, leading to tens of thousands of pixels at the bottom.
	}
	\label{fig:scale_space}
\end{figure*}

\subsection{Scale-aware Generative Models and Infinite Images}
\label{sec:related_generative}

The research on synthesizing infinite images and multi-scale images starts with textures.
Early works on texture synthesis employed non-parametric methods to generate infinite \cite{efros1999texture,efros2001image} and multi-scale \cite{han2008multiscale} textures.
Subsequent advancements revealed that matching statistics of a pretrained CNN increases quality \cite{snelgrove2017high}.

Since the invention of GANs \cite{goodfellow2014generative}, there has been a remarkable surge in the quality of natural image synthesis, with the StyleGAN family being a representative example~\cite{karras2019style,karras2020analyzing,karras2020training,karras2021alias,sauer2022stylegan}.
The success of GANs has motivated researchers to explore the synthesis of very high-resolution or infinite images.
A number of works have modified the GAN pipeline to produce a high-resolution image \cite{lin2021infinitygan,lin2023infinicity,fruhstuck2019tilegan,rodriguezpardo2022SeamlessGAN,zhu2021seamless}.
Apart from GANs, high-resolution synthesis based on transformers  \cite{esser2021taming,wu_nuvainfinity_2022} and diffusion models \cite{zhange2023diffcollage,bond2023infty,lee2023syncdiffusion} have been studied.
However, all these approaches only consider images at a single scale but do not study how to learn and synthesize multi-scale images.

\NEW{
Even within a \emph{single} natural image, content re-appears at multiple scales \cite{glasner2009super,zontak2011internal}.
This property has been exploited to perform image deblurring and super-resolution \cite{michaeli2014blind,ZSSR,bell2019blind}, and to synthesize images with new layouts, structures, and sizes \cite{shaham2019singan,shocher2019ingan,zhou2018non}.
This class of methods can only operate on a narrow range of scales, as self-similarities typically do not persist across orders-of-magnitude scale ranges.
For example, when viewing a large painting from a distance, overall patterns and statistics are quite different from the individual paint strokes and cracks seen up close (Fig.~\ref{fig:teaser}).
Our approach can also be used to reconstruct a single scale space, but assumes unstructured patches at multiple scales as input and therefore does not have to rely only on self-similarity.
}

The works most related to ours are AnyresGAN \cite{chai2022any} and ScaleParty \cite{ntavelis2022arbitrary}, which can handle multi-scale images in both training and inference.
However, these methods consider images of the same semantic level (\eg images of human faces or animals) with a relatively narrow scale range, with the finest scale being only a 4-8x zoom of the coarsest scale.
In this work, we consider a much broader scale range supporting zoom levels up to 256x, which involves the emergence of semantically new content (\eg from an entire galaxy to individual stars) and thus introduces significant challenges.
A concurrent work also studies such drastic multi-scale image synthesis based on diffusion models \cite{wang2023generativepowers}.
However, it requires carefully crafted text prompts for each scale, which is prone to imprecise descriptions.
In addition, this method can only zoom into the center of the image, while our method \NEW{creates full scale spaces.}

\section{Multiscale Images}

The central object of interest in this work is the continuous scale space \scalespace of an image~\cite{iijima1959basic,witkin1987scale}, which captures versions of that image with different upper bounds on frequency content.
We write
\begin{equation}
    \scalespace(\coord, \bandwidth) = \outcolor,
\end{equation}
where $\outcolor \in \mathds{R}^3$ is an RGB color,
$\coord = (x,y)^T \in [-0.5,0.5]^2$ is a continuous location in the image plane,
and $\bandwidth \in \mathds{R}_+$ is an upper frequency limit, also referred to as bandwidth.
A low value of \bandwidth corresponds to an image with only little detail, while increasing \bandwidth progressively reveals structures of higher frequencies~(\refFig{scale_space}a).

We are concerned with scale spaces in which \bandwidth spans a substantial interval
$\left[ \minfrequency, \maxfrequency \right]$.
Here, \minfrequency is already high enough to represent an ordinary image (top face of the cube in \refFig{scale_space}a), and \maxfrequency is orders of magnitude larger than \minfrequency.
To conveniently handle this large dynamic range, we introduce the notion of \emph{scale} \scale, which we define in the logarithmic domain: 
\begin{equation}
    \scale
    =
    \scale(\bandwidth)
    = 
    \log_2
    \left(
        \frac{\bandwidth}{\minfrequency}
    \right)
    \in \mathds{R}_{\geq 0}.
\end{equation}
We further define 
$\maxscale = \scale(\maxfrequency)$ to denote the full dynamic range of scales for a particular \scalespace.
In this work, a typical dynamic range is $\maxscale = 8$, \ie the most detailed image contains frequencies that are 256x higher than those of the coarsest image.

We consider both location \coord and scale \scale continuous parameters.
However, in order to learn a scale-space representation from ordinary images, \ie pixel arrays, as we set out to do in this work, we need to be able to handle discretizations in \coord.
We denote coordinate samples on a regular grid as \discretecoord.
According to the Nyquist-Shannon sampling theorem~\cite{antoniou2006digital}, a signal with bandwidth \bandwidth needs to be sampled at a rate of at least $2\bandwidth$ to capture all available detail and to avoid aliasing, referred to as the Nyquist limit.
In our setting, consequently, the required spatial resolution increases with scale~(white grids in \refFig{scale_space}b).
Specifically, we need images with $\resolution \times \resolution$ pixels, where
\begin{equation}
    \resolution
    =
    \resolution(\scale)
    =
    \lceil \sqrt{2} \minfrequency 2^{\scale + 1} \rceil.
\end{equation}
The factor $\sqrt{2}$ accounts for frequency content along the diagonal of the image plane, where the effective sampling rate is lower.
Given the very high dynamic ranges \maxscale we consider, the required resolution at the finest scale 
$\maxresolution = \resolution(\maxscale)$ 
quickly leads to gigapixel images.
While, in theory, these ultra-high-resolution images constitute perfect data to create a scale-space representation \scalespace, they are extremely difficult to produce.
In practice, a large number of ordinary images is captured using a sophisticated camera setup and stitched together in a post-process~\cite{brady2012multiscale,kopf2007capturing,cossairt2011gigapixel}.
In the next section, we describe a novel, fundamentally different approach for learning a scale space based on adversarial training.

\section{Method}

We propose an algorithm to obtain an orders-of-magnitude scale space \scalespace, which relies on an \emph{unstructured} collection of ordinary, low-resolution images that constitute patch samples of \scalespace.
Each patch has a fixed resolution of $\patchresolution \times \patchresolution$ pixels, slices the scale-space volume at a continuous scale \patchscale, and is centered at a continuous location \patchcoord.
The combination of fixed resolution \patchresolution and varying scale \patchscale across patches leads to varying effective patch sizes in the spatial domain~(green bars in \refFig{scale_space}b):
A patch at a low \patchscale occupies a significant portion of the spatial domain, while a patch at a high \patchscale only covers a tiny fraction of it.
Crucially, \emph{we do not assume any knowledge about the 2D location \patchcoord of each patch}.
This significantly lifts the capture burden, as neither a specialized device nor a sophisticated acquisition protocol is required.
In fact, we will demonstrate that our approach generates high-quality scale spaces even when applied to unstructured image collections from the internet that do not depict the same scene.
We require, however, a coarse estimate of scale \patchscale per patch.
We argue this does not impose unreasonable restrictions, as, for many application domains, \patchscale can be estimated from image metadata, \eg focal length.

\begin{figure}
	\includegraphics[width=0.99\linewidth]{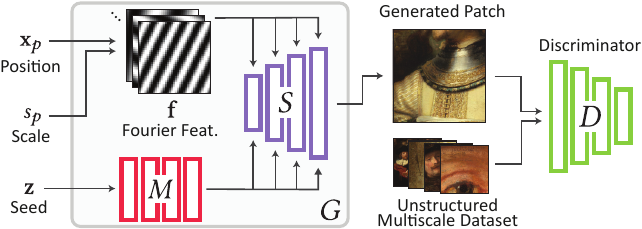}
	\caption{
	Overview of our approach.
    Our multiscale generator \generator takes a patch location \patchcoord and scale \patchscale, as well as a random seed \latentcode as input and synthesizes a corresponding image.
    A discriminator \discriminator compares the distributions of synthesized and data patches.
    Our generator architecture augments an alias-free StyleGAN with carefully designed Fourier features that are distributed across network layers, which allows to synthesize image patches from continuous orders-of-magnitude scale spaces. \NEW{Dataset patches courtesy of \citet{rembrandt}.}
	}
	\label{fig:overview}
\end{figure}

Instead of relying on the common approach of patch alignment and stitching, we train a deep generative model for obtaining \scalespace from the input data~(\refFig{overview}).
Specifically, we design a generator \generator that is able to learn a continuous orders-of-magnitude scale space \generatedscalespace.
\generator takes as input a random vector \latentcode, as well as a continuous patch location \patchcoord and scale \patchscale, and renders one $\patchresolution \times \patchresolution$ image at a time.
Continuously varying \patchcoord and \patchscale produces corresponding slices of the learned \generatedscalespace~\cite{bora2018ambientgan}.
This allows interactive exploration of \generatedscalespace, while adjacent synthesized patches can be stitched together to yield seamless, arbitrary-resolution composites.
We train \generator in an adversarial fashion~\cite{goodfellow2014generative}, \ie we jointly train a discriminator \discriminator that compares the distribution of generated patches from \generatedscalespace to the distribution of data patches from \scalespace.
\NEW{During training, \latentcode, \patchcoord and \patchscale are randomly sampled.}
An additional consistency loss~\cite{chai2022any,ntavelis2022arbitrary} encourages coherency across scales.

Our statistical approach is surprisingly versatile and supports two modes of operation.
First, given a collection of unstructured input patches from a \emph{single} scale space \scalespace, \ie multiscale observations of the \emph{same scene}, \generatedscalespace converges to a plausible, coherent approximation of \scalespace.
While we find providing random inputs \latentcode to \generator is necessary for stable training, a converged generator \generator disregards \latentcode, producing negligible variations of the output.
We refer to this solution as \emph{pseudo-reconstruction}, as \generatedscalespace might differ from \scalespace in the arrangement of details, but captures the overall multiscale structure well.
Note that in addition to missing knowledge about the patch locations \patchcoord, the patches do not exhaustively cover \scalespace at all scales.
However, due to the generative capabilities of our framework, missing content will be seamlessly and consistently hallucinated across all scales.
Along with implicitly performing an alignment of the input patches, \generator is a very compact representation:
It has up to 885x less parameters than the corresponding gigapixel image at \maxscale requires RGB values stored in its pixel grid.

In the second mode of operation, we train \generator with a collection of input patches from \emph{different} environments.
The training patches depict different scenes from a (narrow) class at different scales.
In this case, a converged \generator produces a \emph{distribution of scale spaces} $p(\generatedscalespace)$, where different \latentcode result in different samples from that distribution.

To achieve our goal of learning images across scales using adversarial training, we require two essential ingredients:
First, we design a generator \generator that can synthesize scale spaces \generatedscalespace with large dynamic range \maxscale.
Second, we develop a training procedure that allows \generator to \emph{robustly} learn \emph{coherent} scale spaces.
We give details on these ingredients in \refSec{architecture} and \refSec{training}, respectively.

\subsection{Multiscale Generator}
\label{sec:architecture}

We require a generator \generator that is able to encode orders-of-magnitude scale spaces \generatedscalespace.
Importantly, while the output of \generator is a pixel grid, the model has to be intrinsically continuous to be a faithful representation of \scalespace and to allow for arbitrary translation and zooming.
Fortunately, a continuous generator design is available in the form of alias-free StyleGAN (StyleGAN3)~\cite{karras2021alias}, which allows continuous translation of the generated content and has been shown to support some zooming~\cite{chai2022any}.
However, we find that a vanilla StyleGAN3 generator architecture is not able to synthesize scale spaces for the high \maxscale we require.
Therefore, we extend it to the multiscale setting.

\begin{figure}
	\includegraphics[width=0.9\linewidth]{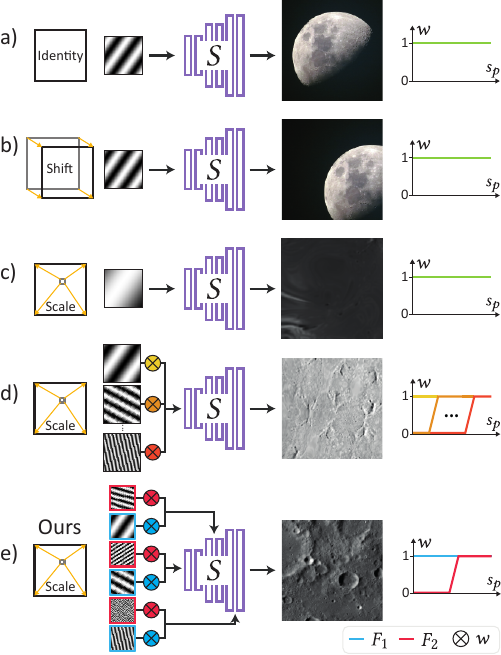}
	\caption{
    \emph{(a)} The StyleGAN3 generator rasterizes Fourier features \fourierfeature (one is shown) and feeds them through a synthesis network \synthesisnetwork to obtain an output image.
    \emph{(b)} A spatial offset of \fourierfeature results in a shifted image.
    \emph{(c)} Scaling up \fourierfeature leads to flat feature maps, which \synthesisnetwork cannot translate into a meaningful image.
    \NEW{In the setups \emph{a)-c)}, the features \fourierfeature are not modulated (constant weighting \weightfct).}
    \emph{(d)} Progressively blending in different $\fourierfeature$ using the weighting function $\weightfct(\patchscale)$ results in a permanent re-scaling of individual features \fourierfeature \NEW{(three out of many are shown)}, leading to unstable results.
    \emph{(e)} We create Fourier features in bins (\NEW{two bins -- pink and blue -- are shown)} and blend in all features per bin at the same time. \NEW{This leads to a significant reduction of blending (here, only the pink bin needs blending).} Additionally, we inject features into different layers of \synthesisnetwork, significantly enhancing coherency across scales.
	}
	\label{fig:frequency_blending}
\end{figure}

The StyleGAN3 generator relies on Fourier features 
$\fourierfeature \in \mathds{R}^\fourierfeaturedim$,
\ie directional 2D sinusoids evaluated on a regular grid \discretecoord, \NEW{\ie they can be represented in the spatial domain as a pixel grid with \fourierfeaturedim channels} (\refFig{frequency_blending}a):
\begin{equation}
\label{eq:fourier_eval}
    \fourierfeature_j(\discretecoord)
    =
    \sin 
    \left(
        2 \pi \fourierfeaturefreq_j^T \discretecoord
    \right),
\end{equation}
where $\fourierfeaturefreq_j \in \mathds{R}^2$ are \fourierfeaturedim different frequency vectors.
The features \fourierfeature are fed into a sequence of layers of a synthesis network \synthesisnetwork, each of which performs non-linear operations.
Occasionally, intermediate neural features are up-sampled to a higher spatial resolution.
The non-linear operations are modulated by ``style'' vectors, which arise from feeding the latent code \latentcode through a mapping network \mappingnetwork (\refFig{overview}).
Both, non-linear operations and up-sampling, are carefully designed such that the resulting neural features only contain spatial frequencies below the Nyquist limit dictated by the resolution of the respective layer.
A direct consequence of this approach is that the entire generator can be treated as a continuous function, despite relying on regular grids for the actual computations.
We choose a variant of the generator, where the neural operations are applied point-wise (R-configuration)~\cite{karras2021alias}.
In addition to obtaining rotational equivariance, this configuration is well suited for multiscale generation, as the alternative -- spatial convolutions -- typically operates on fixed-size neighborhoods, whose meaning varies with scale.
We synthesize images of resolution $\patchresolution=256$.

By design, procedurally shifting Fourier features \fourierfeature at the beginning of the processing sequence results in alias-free translation of the output image (\refFig{frequency_blending}b).
To incorporate scaling, in the first step, we transform all grid coordinates \discretecoord via
\begin{equation}
\label{eq:coord_scaling}
    \transform(\discretecoord; \patchcoord, \patchscale)
    =
    2^{-\patchscale} 
    \left(
        \discretecoord - \patchcoord
    \right),
\end{equation}
and feed the corresponding shifted and scaled Fourier features
$
\fourierfeature_j
\left(
    \transform ( \discretecoord; \patchcoord, \patchscale)    
\right)
$
into \synthesisnetwork.
Unfortunately, choosing a high \patchscale stretches \fourierfeature to such an extent that it degrades to an almost constant function (\refFig{frequency_blending}c).
Unsurprisingly, we find that \synthesisnetwork cannot learn to synthesize meaningful images given such an input.
Simply increasing frequencies 
$\| \fourierfeaturefreq_j \|$
is not a solution to the problem, as we need to stay below the Nyquist limit of the first layer.
Therefore, in a second design iteration, we could indeed sample higher frequencies $\fourierfeaturefreq_j$, but progressively blend them in only after \transform has stretched out the corresponding $\fourierfeature_j$ far enough to stay below the Nyquist limit, using some blending function \weightfct (\refFig{frequency_blending}d).
While this strategy provides \synthesisnetwork with meaningful frequency content across all scales, we observe drifting and tearing in the output images during zooming.
This is because we are interfering with the positional encoding provided by the Fourier features in \refEq{fourier_eval}:
The effect of scaling $\fourierfeature_j$ is not distinguishable from the effect of shifting the input position \coord.
This ambiguity is exacerbated by the severe non-linearity of \synthesisnetwork.

\begin{figure}
	\includegraphics[width=1\linewidth]{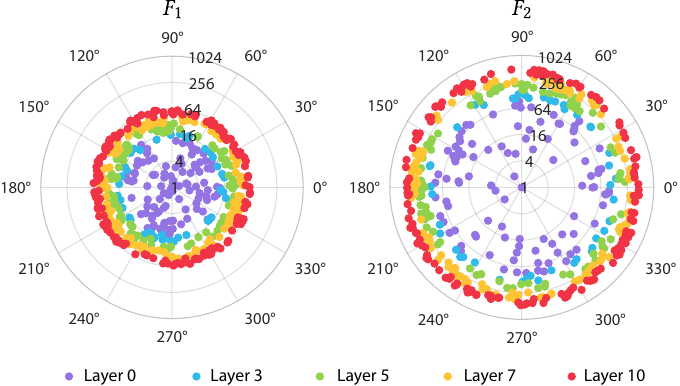}
	\caption{
        \NEW{Distribution of Fourier features $\fourierfeature_j$ across bins and generator layers. Each point represents a 2D frequency $\fourierfeaturefreq_j$ (magnitude is visualized in log space). Frequencies for different bins $\freqset_1$ and $\freqset_2$ are displayed separately.
        Colors signify the layer of the synthesis network into which a frequency is injected -- the higher the frequency, the later the injection.
    }
	}
	\label{fig:frequency_distribution}
\end{figure}

We address this problem in our final design (\refFig{frequency_blending}e), which is based on two crucial observations.
First, careful binning of Fourier features and simultaneous blending per bin significantly reduces the amount of re-weighting necessary.
Second, the less non-linear layers are operating in between Fourier features and the final image, the less positional distortions they can cause.
Consequently, we employ a re-assignment of binned Fourier features to different layers of \synthesisnetwork.

For frequency binning, we consider non-overlapping scale intervals of size 
$\Delta\scale$, 
and create 
$\setcount = \lceil \sfrac{\maxscale}{\Delta\scale} \rceil$
corresponding frequency bins $\freqset_k$, where 
$k \in \{0 \ldots \setcount-1 \}$.
For each $\freqset_k$, we randomly sample 512 frequencies with a maximum magnitude of 
$2^{\basescale + \Delta\scale \cdot k}$, where \basescale is a hyper-parameter. \NEW{\refFig{frequency_distribution} shows a frequency distribution for two bins}.
We set $\Delta\scale=3$ and $\basescale=6$ in all our experiments.
We now define a weighting function that blends in all Fourier features per bin $\freqset_k$ as a function of scale \patchscale:
\begin{equation}
\label{eq:weighting}
    \weightfct_k(\patchscale)
    =
    \min
    \left(
        1, 
        \max
        \left(
            0,
            \patchscale - \Delta\scale \cdot k + 1
        \right)
    \right).
\end{equation}
This weighting scheme blends in entire bins of Fourier features simultaneously across regularly spaced, narrow intervals, which significantly reduces the amount of blending happening during zooming.

We inject the so-obtained weighted Fourier features into different layers of \synthesisnetwork~\cite{diolatzis2023mesogan}.
Specifically, in addition to injection into the first layer, we concatenate Fourier features $\fourierfeature_j$ to neural features after each up-sampling layer.
The assignment of $\fourierfeature_j$ to the individual layers is based on per-layer Nyquist limits.
We iterate over all injection layers in order and assign to the current layer those $\fourierfeature_j$ that have not yet been assigned, if the following condition is met:
At the scale \patchscale where the feature is blended in completely via \refEq{weighting}, the scaling in \refEq{coord_scaling} lets $\fourierfeature_j$ fall below the layer's Nyquist limit \NEW{(color coding in \refFig{frequency_distribution})}.

Our generator is now able to synthesize \emph{detailed} and \emph{coherent} content across orders-of-magnitude scale ranges.
Let's train it!

\subsection{Training}
\label{sec:training}

Our multiscale generator is trained in an adversarial fashion~\cite{goodfellow2014generative} using an image discriminator \discriminator that compares the distribution of generated patches to the distribution of training patches (\refFig{overview}).
\NEW{In addition to sampling the random vector \latentcode, we also randomly sample the patch location \patchcoord and scale \patchscale.}
Regarding training setup and hyper-parameters, we largely follow the official StyleGAN3~\cite{karras2021alias} implementation and corresponding recommendations of the authors, involving R1 regularization~\cite{mescheder2018training} and adaptive discriminator augmentation~\cite{karras2020training}.
However, we find that our multiscale setting poses two significant challenges:
\emph{(i)} training stability, and \emph{(ii)} consistency of the learned \generatedscalespace across scales.
We address these items using a progressive patch sampling scheme and a scale consistency loss, respectively.

\subsubsection{Progressive Patch Sampling}
\begin{figure}
	\includegraphics[width=\linewidth]{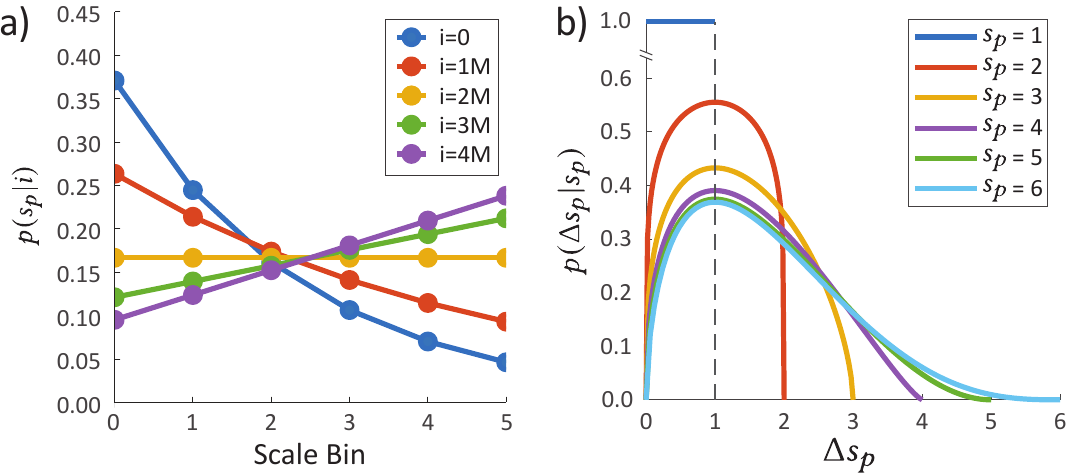}
	\caption{
    (\emph{a}) Sampling distribution of scale bins during training, where $i$ denotes the number of images used so far.
    (\emph{b}) Sampling distributions of scale offset $\Delta\patchscale$ for different patch scales $\patchscale$. 
    }
	\label{fig:patch_sampling}
\end{figure}

We observe that simple uniform random sampling of patches across all scales does not converge to satisfactory results~\cite{chai2022any}.
We therefore use a patch sampling scheme that prioritizes coarse scales at the beginning of training and progressively shifts attention towards the finer scales (\refFig{patch_sampling}a).
This happens both for generated and for data patches; recall that we assume access to a coarse estimate of scale \patchscale per data patch.
We operate on scale bins of width one to avoid relying on exact scale labels in the data.
Specifically, we transition from a negative exponential distribution (blue curve in \refFig{patch_sampling}a), over a uniform distribution (yellow curve), to a linearly increasing distribution (pink curve).
We find that this progressive strategy leads to significantly improved training stability.

\subsubsection{Scale Consistency Loss}
\label{sec:consistency_loss}
To encourage scale-coherent \generatedscalespace, we follow ideas from \citet{chai2022any} and add an additional loss term that compares two generated patches, separated by a scale offset $\Delta\patchscale \in \mathds{R}_+$~\cite{irani1991improving}:
\begin{equation}
\label{eq:loss}
    \mathcal{L}_\textrm{\scale}
    =
    \mathbb E_{\latentcode, \patchcoord, \patchscale, \Delta\patchscale}
    \left[
        \patchdistance
        \left(
            \resampling_{\Delta\patchscale} \left(
                \generator(\latentcode, \patchcoord, \patchscale)
            \right),
            \generator(\latentcode, \patchcoord, \patchscale - \Delta\patchscale)
        \right)
    \right].
\end{equation}
Here, $\resampling_{\Delta\patchscale}$ is a function that downsamples an image by a factor $2^{\Delta\patchscale}$.
\patchdistance is an image distance metric that we apply to the two generated patches, only considering pixels that appear in \emph{both} patches.
We implement \patchdistance using a linear combination of $\ell_1$-norm and LPIPS~\cite{zhang2018perceptual} distance.
\NEW{
While we could backpropagate gradients through both generator instances in \refEq{loss}, we find that training stability improves when we randomly choose only one of the generator instances to receive gradients in each training iteration.
}

While we uniformly sample \patchcoord in \refEq{loss}, we find that the choice of sampling distribution for $\Delta\patchscale$ has a strong influence on the scale consistency of \generatedscalespace for our large \maxscale.
To make sure \generatedscalespace is globally consistent, we want $\Delta\patchscale$ to be sampled in the full range $[0, \patchscale]$, \ie every patch is compared against arbitrarily zoomed-out counterparts up to $\patchscale=0$.
However, a uniform distribution over $[0, \patchscale]$ is not an optimal choice.
On the one hand, sampling a very low value for $\Delta\patchscale$ results in almost identical patches, wasting training resources.
On the other hand, sampling a very high value results in images of significantly different relative resolutions, \ie $\resampling_{\Delta\patchscale}$ produces very low-resolution patches to be compared against, providing not much of a supervision signal either.

Our solution relies on the beta distribution B, which accounts for all the above considerations:
\begin{equation}
    p(\Delta\patchscale | \patchscale) 
    = 
    \frac
    {B(\frac{\Delta\patchscale}{\patchscale}; \alpha,\beta)}
    {\patchscale},
\end{equation}
where
$\alpha = \sqrt[4]{\max(1, \patchscale)}$, and
$\beta = (\alpha - 1) \max(1, \patchscale) - \alpha + 2$.
The parameters are chosen such that $p(\Delta\patchscale | \patchscale)$ has its mode at $\Delta\patchscale=1$ and gradually falls off in both directions, while still covering the entire available scale interval (\refFig{patch_sampling}b).

\section{Evaluation}
\label{sec:evaluation}

We evaluate our approach on the tasks of multiscale pseudo-recon- struction (\refSec{reconstruction}) and generation (\refSec{generation}).
We further analyze the components and properties of our method (\refSec{analysis}).
We urge the reader to watch our supplemental video, in which we demonstrate continuous zooming and panning through our obtained scale spaces.
Our generator runs at 20\,fps, which allows highly interactive exploration of our scale spaces.

\paragraph{Datasets \NEW{and Training Details}}
We use a total of \NEW{seven} datasets for our evaluation, all containing unstructured patches at a resolution of $\patchresolution=256$.
For \dataset{Himalayas} and \dataset{Spain}, we consider multiscale satellite data~\cite{sentinel}.
Both datasets cover a square geographic region with $\maxscale=8$.
Scale labels are obtained from satellite metadata.
To enable a broader range of quantitative evaluations, we additionally employ three gigapixel images -- \dataset{Milkyway}~\cite{wojczynski2021} ($\maxscale=6$), \dataset{Moon}~\cite{speyerer2011lunar} ($\maxscale=6$) and \dataset{Rembrandt}~\cite{rembrandt} ($\maxscale=8$) --, from which we extract random patches at multiple scales to simulate our input setting.

To evaluate generative capabilities, we only consider patches sampled from the finest four scales of each dataset source, denoted \dataset{HimalayasGen}, \dataset{SpainGen}, \dataset{MilkywayGen}, \dataset{MoonGen}, and \dataset{RembrandtGen}, respectively.
The so-obtained data forces our models to learn scale-space distributions.
Further, the \dataset{Sunflowers} \NEW{and \dataset{Bricks} datasets are} composed of a collection of images from Flickr.
\NEW{For \dataset{Sunflowers}, images were queried using the text strings ``sunflower field'' and ``sunflower'', while for \dataset{Bricks} the search strings were ``brick wall'' and ``bricks and cracks''.}
Obtained images are randomly cropped and down-sampled to our target resolution to achieve $\maxscale=4$.
Coarse scale labels are assigned semi-automatically, taking into account the specific query, image resolution, and crop window size.
Naturally, these datasets contains a variety of different scenes.

Datasets for pseudo-reconstruction contain 96k (\dataset{milkyway} and \dataset{moon}) or 156k patches (\dataset{himalayas}, \dataset{spain}, \dataset{rembrandt}), while those for generation contain 120k images, \NEW{except for \dataset{Sunflowers} (185k patches) and \dataset{Bricks} (234k patches).}
\NEW{The supplemental document lists more detailed dataset statistics.}

\NEW{We obtain converged models after 52-75 hours of training using eight A100 GPUs. Our models occupy 38-62MB of disk space. During inference, they require 2.8-3.1GB of VRAM.}

\begin{table}[]
    \centering
    \caption{Quantitative evaluation of multiscale pseudo-reconstruction.}
    \renewcommand{\tabcolsep}{0.045cm}
    \label{tab:quant_reconstruction}
    
    \begin{tabular}{lcrrrrrrrrrr} 
        \multirow{2}{*}[-0.5ex]{Dataset} & \multirow{2}{*}[-0.5ex]{\maxscale} & \multirow{2}{*}[-0.5ex]{FID $\downarrow$} & \multicolumn{4}{c}{Scale Consistency} & \multirow{2}{*}[-0.5ex]{PSNR\textsubscript{GT}$\uparrow$} \\
        \cmidrule(lr){4-7}
        & & & \footnotesize{Bias $\downarrow$} & \footnotesize{Angle $\downarrow$} & \footnotesize{EMD $\downarrow$} & \footnotesize{PSNR\textsubscript{inter} $\uparrow$} \\
        \midrule

\dataset{Himalayas} & 8 & 19.1 & 0.06 & 1.2 & 1.08 & 19.5 & - \\
\dataset{Spain} & 8 & 8.4 & 0.07 & 1.0 & 0.98 & 23.3 & - \\
\dataset{Milkyway} & 6 & 8.6 & 0.21 & 1.5 & 1.83 & 25.9 & 24.0 \\
\dataset{Moon} & 6 & 9.0 & 0.38 & 1.1 & 1.28 & 28.9 & 25.8 \\
\dataset{Rembrandt} & 8 & 14.6 & 0.20 & 1.2 & 1.91 & 28.9 & - \\
    
        \bottomrule
    \end{tabular}
\end{table}

\begin{table}[]
    \centering
    \caption{Quantitative evaluation of multiscale generation.}
    \renewcommand{\tabcolsep}{0.115cm}
    \label{tab:quant_generation}
    
    \begin{tabular}{llrrrrrrrr} 
        \multirow{2}{*}[-0.5ex]{Dataset} & \multirow{2}{*}[-0.5ex]{Method} & \multirow{2}{*}[-0.5ex]{FID $\downarrow$} & \multicolumn{3}{c}{Scale Consistency} \\
        \cmidrule(lr){4-6}
        & & & \footnotesize{Bias $\downarrow$} & \footnotesize{Angle $\downarrow$} & \footnotesize{EMD $\downarrow$} \\
        \midrule
\multirow{3}{*}{\dataset{MilkywayGen}} & AnyresGAN & 30.6 & 0.81 & 11.3 & 12.22
 \\& PULSE & - &  0.43 & 6.9 & 8.69 \\
 & \textbf{Ours} & 28.3 & 0.11 & 1.2  & 1.55 \\
 \midrule
\multirow{3}{*}{\dataset{MoonGen}} & AnyresGAN & 18.4 & 0.95 & 11.8 & 12.8 \\
& PULSE & - & 0.41 & 19.7 & 17.75 \\
 & \textbf{Ours} & 6.3 & 0.23 & 2.2 & 2.73 \\
    \midrule
\dataset{HimalayasGen} & \textbf{Ours} & 19.4 & 0.11 & 1.1 & 1.28 \\
\dataset{SpainGen} & \textbf{Ours} & 9.7 & 0.05 & 1.1 & 1.17 \\
\dataset{RembrandtGen} & \textbf{Ours} & 18.9 & 0.36 & 3.7 & 7.37 \\
\dataset{Sunflowers} & \textbf{Ours} & 9.8 & 0.09 & 1.1 & 1.24 \\
\dataset{\NEW{Bricks}} & \textbf{\NEW{Ours}} & \NEW{6.8} & \NEW{0.08} & \NEW{1.03} & \NEW{1.16} \\
    
        \bottomrule
    \end{tabular}
\end{table}

\paragraph{Evaluating Scale Consistency}

One crucial property of a scale space is its consistency across scales.
We employ two procedures to quantify this.
First, we create sequences of images, gradually zooming in. Using off-the-shelf optical flow estimation~\cite{teed2020raft}, we compute per-pixel motion trajectories (\refFig{trajectory_comparison}).
As a first metric (inspired by tOF in \cite{chu2020learning}), we average all flow vectors to obtain an estimate of overall bias; a perfect solution has radial trajectories only (\refFig{trajectory_comparison}, right) and, thus, zero bias.
For our second metric, we fit a line to each trajectory and compute the angular difference to the ground-truth trajectory \cite{cogalan2023video}.
As a third metric, we compute the earth mover's distance (EMD)~\cite{rubner1998metric} between each motion trajectory and the ground truth.

In a second procedure, we generate full-resolution scale-space slices at all integer scales.
Notice that this involves spatial resolutions from $256 \times 256$ for $\scale=0$ up to $65\text{k} \times 65\text{k}$ for $\scale=8$.
We now compute the PSNR between all slice pairs ($\text{PSNR}_\text{inter}$), where higher-resolution images are downsampled to match the lower-resolution ones.
We also compute the PSNR with respect to ground truth when it is available ($\text{PSNR}_\text{GT}$).
As reconstructed scale spaces do not exactly align with the references, we perform a global alignment using translation and isotropic scaling.

\subsection{Multiscale Pseudo-Reconstruction}
\label{sec:reconstruction}

We show scale-space pseudo-reconstructions in \refFig{results}, top and a corresponding quantitative evaluation in \refTab{quant_reconstruction}.
More results can be found in the supplemental video.
We observe that we are able to successfully learn orders-of-magnitude scale spaces, which allow coherent zooming into any location.
The last column in \refTab{quant_reconstruction} reveals that our reconstructions are quite close to the ground truth on average.
We investigate this further in \refFig{metric_details}a, where we break down reconstruction accuracy per scale.
We see that accuracy decreases for higher scales.
This is because our model has the freedom to hallucinate high-frequency content as long as the overall structure is coherent
\NEW{\ie our approach is a form of generative compression~\cite{santurkar2018generative}}.
We study this behavior in an additional experiment, where we intentionally filter out training patches that overlap a certain spatial region.
In \refFig{exclusion_comparison}, we show that this exclusion leads to inaccurate yet highly plausible content.
In the supplemental document, we show a best-effort result of stitching a subset of our training patches using Adobe Photoshop.

\subsection{Multiscale Generation}
\label{sec:generation}

\begin{figure}
    \centering
    \includegraphics[width=\linewidth]{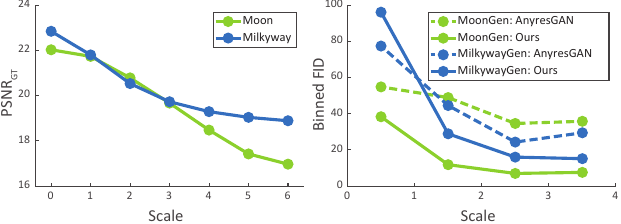}
    \caption{
        (\emph{a}) Reconstruction accuracy of our approach as a function of scale.
        (\emph{b}) FID scores as a function of scale in the generative setting.
    }
    \label{fig:metric_details}
\end{figure}

\begin{figure*}
    \centering
    \includegraphics[width=0.98\linewidth]{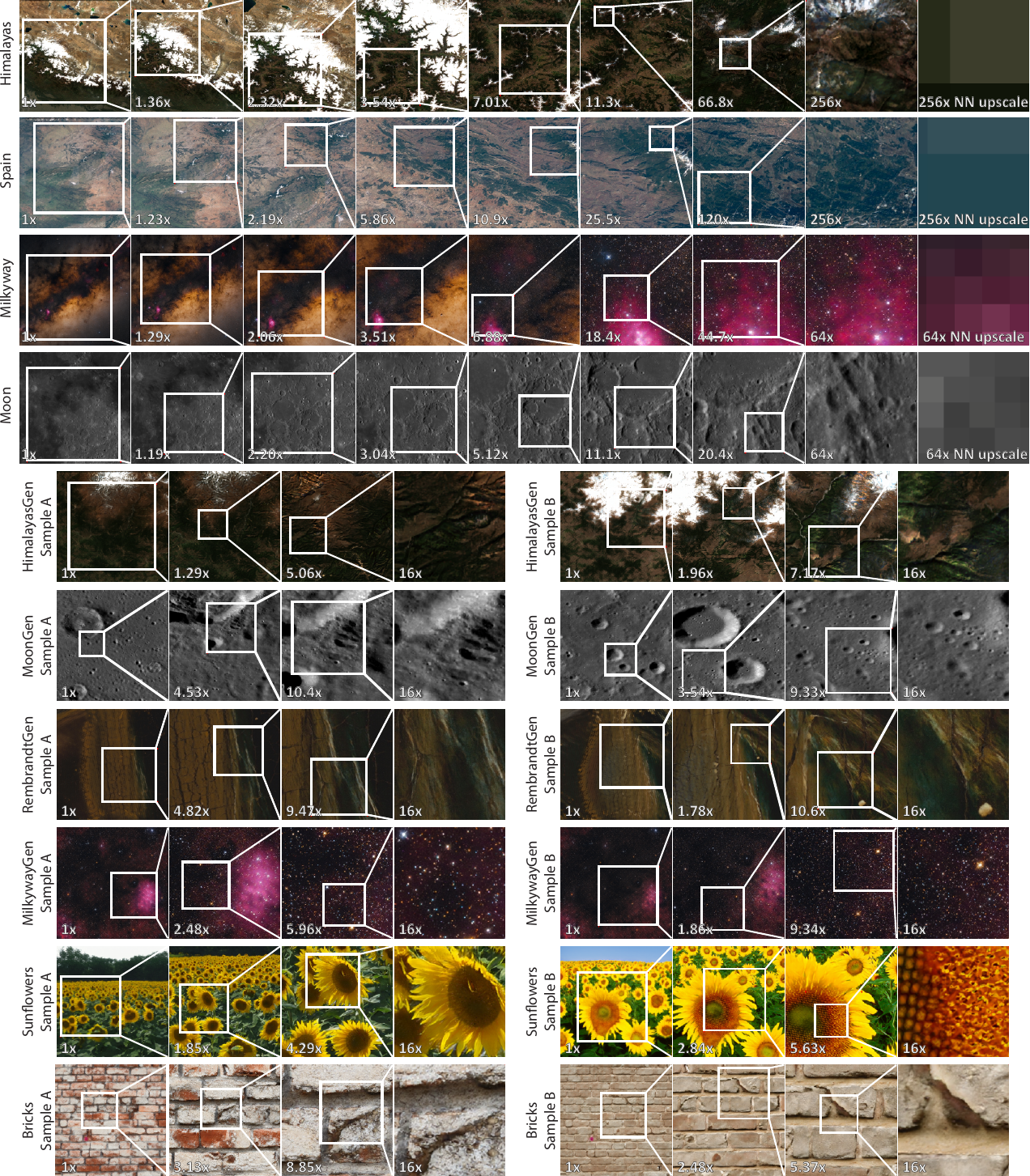}
    \caption{
    Traversal of our scale spaces. 
    The top four rows show pseudo-reconstructions, while the bottom six rows demonstrate generative scale spaces.
    The last column in the top four rows shows the upsampled version of the image in the first column that corresponds to the area that the image in the second-to-last column covers.
    Please refer to our supplemental video for demonstrations of continuous zooming and panning.}
    \label{fig:results}
\end{figure*}

\begin{figure*}
    \centering
    \includegraphics[width=0.97\textwidth]{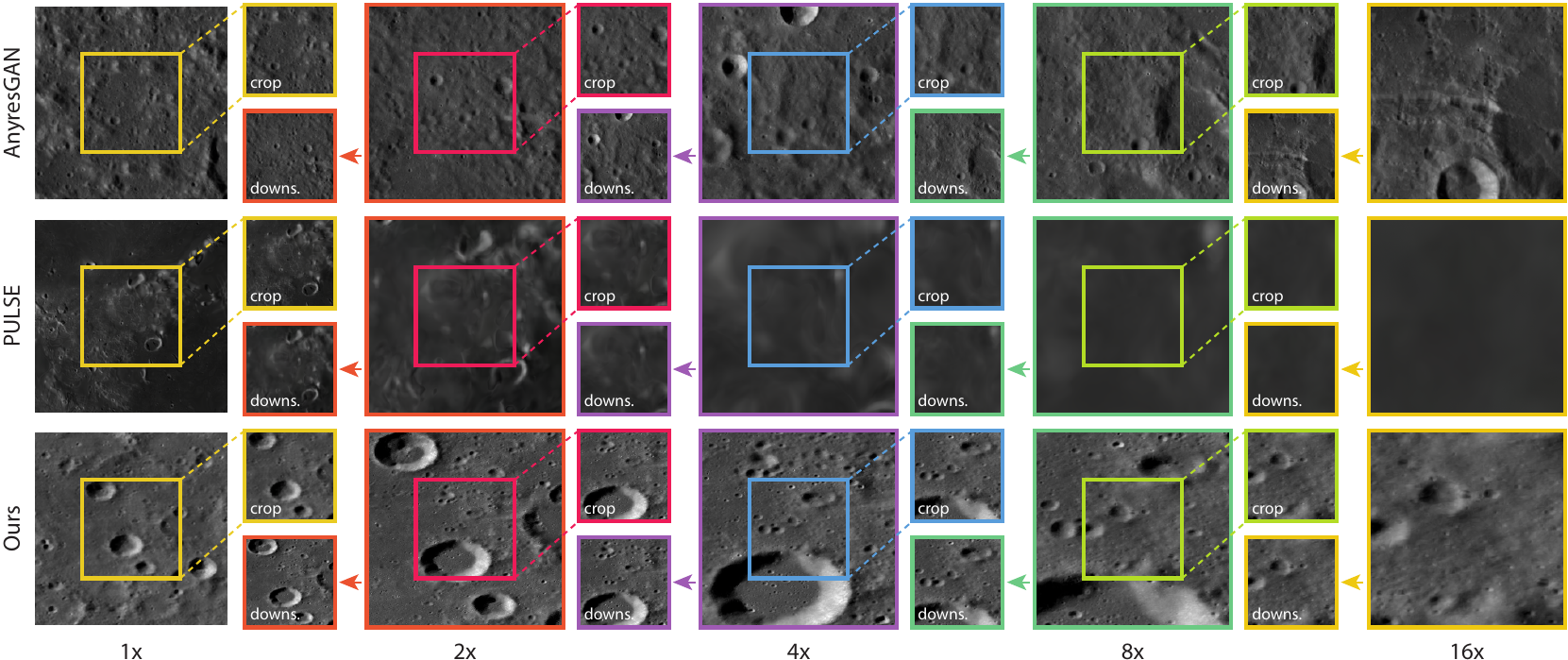}
    \caption{\NEW{Qualitative comparison of different methods (rows).
    Every column (large images) depicts a 2x zoom into the center of the image in the previous column.
    The small insets compare the same image region across two adjacent scales (the coarser scale is cropped, the finer scale is downsampled).
    Please note the lack of scale consistency for AnyresGAN and the loss of details for PULSE. 
    Only our method is consistent and produces rich details across scales.}}
    \label{fig:SOTA_comparison}
\end{figure*}

\begin{figure*}
    \centering
    \includegraphics[width=\textwidth]{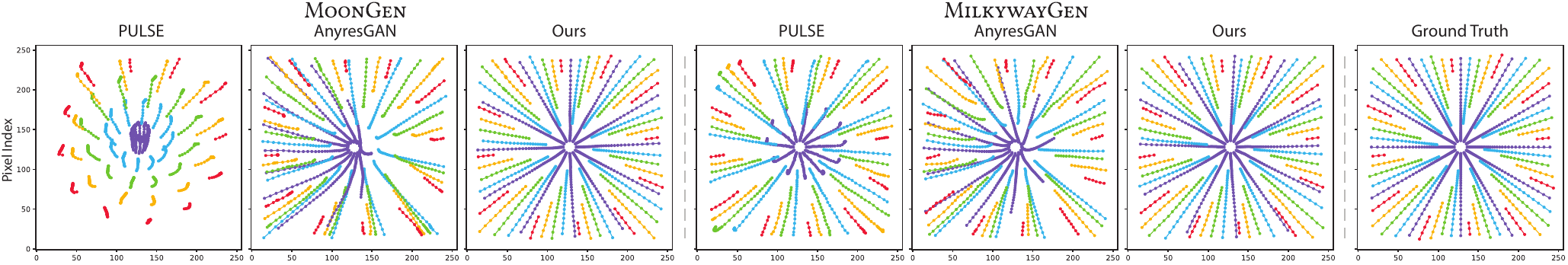}
    \caption{Pixel trajectories for representative samples of different methods. 
    \NEW{
    To obtain the trajectories, we compute an image sequence zooming into the image center and concatenate optical flow vectors~\cite{teed2020raft} of adjacent images in the sequence per pixel. Here, we visualize the trajectories for 60 pixels only to avoid clutter. A perfect flow field is radial, \ie linearly expanding from the center (Ground Truth, right).
    Compared to the baselines, our method provides trajectories closest to the ground truth. 
    }
    }
    \label{fig:trajectory_comparison}
\end{figure*}

Results on multiscale generation are shown in \refFig{results}, bottom and \refTab{quant_generation}.
We compare against two baselines, AnyresGAN~\cite{chai2022any} and PULSE~\cite{menon2020pulse} on two datasets.
Details on how we modify these baselines to be able to handle our setting are provided in the supplemental document.
We observe that our scale spaces are of significantly higher quality than those of the baselines, both in terms of patch distributions measured using FID~\cite{heusel2017gans} and scale consistency.
In \refFig{metric_details}b, we break down FID scores into scale bins, revealing that slices of our scale-space samples are well-behaved across scales.
Notice that FID scores for coarse scales are less reliable due to less available data.
In \NEW{\refFig{SOTA_comparison} we demonstrate qualitative results, while} \refFig{trajectory_comparison} shows flow trajectories of representative samples across methods, confirming that our approach delivers highly scale-consistent results.

\subsection{Analysis}
\label{sec:analysis}

\begin{figure}
    \centering    \includegraphics[width=\linewidth]{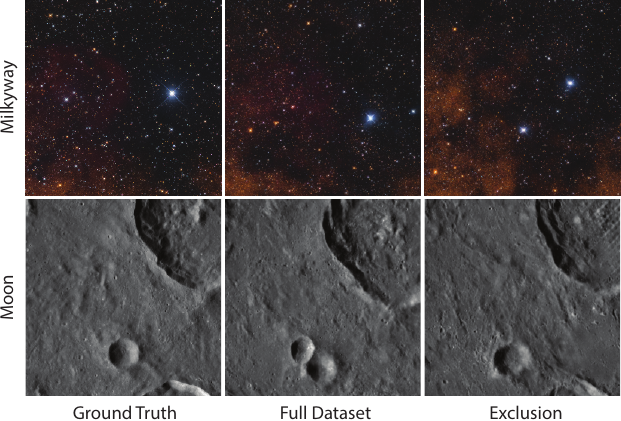}
    \caption{
        Pseudo-reconstruction of an area not present in the training dataset.
        Our approach synthesizes plausible and coherent details, which do not necessarily match the reference, \eg stars are placed at different locations. \NEW{Milkyway image courtesy of Bartosz Wojczyński \shortcite{wojczynski2021}. }
    }
    \label{fig:exclusion_comparison}
\end{figure}

\paragraph{\NEW{Compression}}
Our model requires 14M parameters, which is more compact than the models used in AnyresGAN (32M parameters)  and PULSE (18M parameters).
\NEW{
In contrast, an RGB gigapixel image at a corresponding resolution of $65\text{k}\times 65\text{k}$ pixels requires 13B scalars to be stored.
Thus, in terms of raw parameter reduction, our approach achieves a compression of 885x.
Lossless or lossy compression can be applied on top of both approaches, e.g., JPEG for images and model weight compression for StyleGAN \cite{belousov2021mobilestylegan}.
To shed some light on practical compression capabilities, we JPEG-compress the raw \dataset{Milkyway} gigapixel image to obtain a file size equal to our \emph{uncompressed} model (JPEG quality: 32) and measure image quality at the finest scale. As our model only performs a pseudo-reconstruction in which details do not align with the reference (\refFig{exclusion_comparison}), pixel-wise PSNR is not an expressive metric for this task. We instead opt for patch-based FID, which yields a score of 44 for our model and 114 for the JPEG image, indicating that storing pixels of gigapixel images is inferior to our continuous generative approach.
}

\paragraph{\NEW{Ablations}}
In \refTab{ablations}, we report the results of ablation studies on the pseudo-reconstruction task using \dataset{Milkyway}.

We first consider alternatives to our Beta sampling in \refSec{consistency_loss}.
We compare against a uniform sampling of the full-scale range, as well as two scales.
We observe that our Beta sampling improves all relevant metrics.

\begin{figure}
	\includegraphics[width=0.99\linewidth]{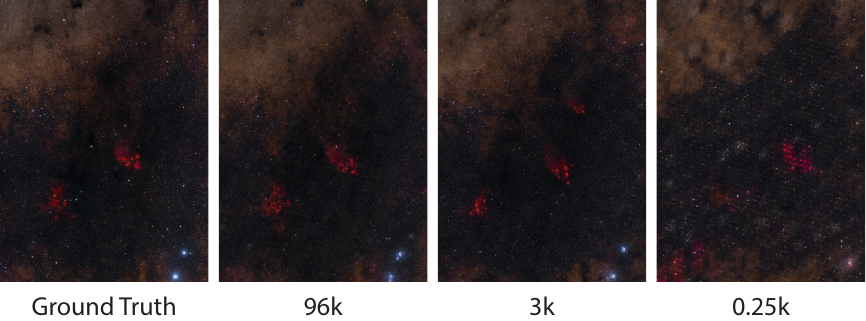}
	\caption{	
        \NEW{Influence of dataset size on reconstruction quality. Structures gradually dissolve into regular patterns as training data becomes more sparse.} \NEW{Ground truth image courtesy of Bartosz Wojczyński \shortcite{wojczynski2021}.}
	}
	\label{fig:density_ablation}
\end{figure}

We further study the effect of reducing dataset size.
We observe that, unsurprisingly, quality and consistency are highest with the full dataset \NEW{containing 96k patches, but even a significant reduction in dataset size does not have a dramatic negative effect on our model.
Interestingly, our method still converges when using only 250 images distributed across the six scales.
This, however, comes with a severe degradation in image quality, while scale consistency improves.
Training on only 100 images diverges.
\refFig{density_ablation} shows corresponding qualitative results.}

\NEW{Finally, we investigate the reliance of our method on exact scale labels by adding uniform random noise with an interval of two scales to the labels.
We observe a minor drop in FID score, while scale consistency metrics are barely affected.
}

\subsection{\NEW{Limitations}}

\begin{figure}
	\includegraphics[width=0.99\linewidth]{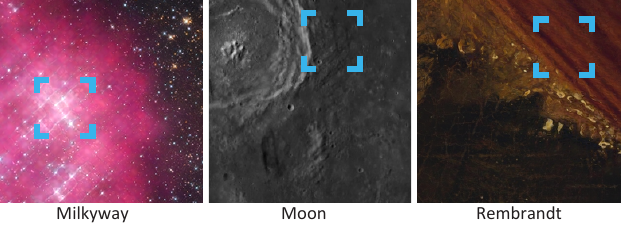}
	\caption{	
        \NEW{Our solution occasionally produces artifacts in the form of overlays with parallel lines in certain sparse areas.}
	}
	\label{fig:artifacts}
\end{figure}

\NEW{
Our distribution of Fourier features across generator layers (\refSec{architecture}) comes with a disadvantage:
Compared to a vanilla setup, the generator network has less capacity to turn procedural frequency content into final image output.
This can occasionally lead to regularity artifacts in the generated patches.
As illustrated in \refFig{artifacts}, some images are faintly overlaid with parallel lines.
We observe that these artifacts mostly appear at the finest scales.
Additionally, we occasionally observe saturated colorful blobs in our generated scale spaces.
The origin of these artifacts can also be traced back to the late injection of Fourier features.
}

\NEW{As with many generative approaches, training times of our method are substantial. The (manual) effort our approach allows to save during data capture has to be paid by increased processing time.
While we are not aware of any other method that is capable of handling the unstructured inputs our approach can process, more research at the foundations of generative modeling are necessary to allow end users with only a single workstation to fully benefit from our technology.
}

\begin{table}[]
    \centering
    \caption{Ablations.}
    \renewcommand{\tabcolsep}{0.05cm}
    \label{tab:ablations}
    
    \begin{tabular}{lrrrrrrrrrr} 
        \multirow{2}{*}[-0.5ex]{Config.} & \multirow{2}{*}[-0.5ex]{FID $\downarrow$} & \multicolumn{4}{c}{Scale Consistency} & \multirow{2}{*}[-0.5ex]{PSNR\textsubscript{GT}$\uparrow$} \\
        \cmidrule(lr){3-6}
        & & \footnotesize{Bias $\downarrow$} & \footnotesize{Angle $\downarrow$} & \footnotesize{EMD $\downarrow$} & \footnotesize{PSNR\textsubscript{inter} $\uparrow$} \\
        \midrule

Uniform\textsubscript{full} & 9.5 & 0.20 & 1.6 & 2.77 & 25.1 & 23.8 \\
Uniform\textsubscript{2} & 14.6 & 0.19 & 1.8 & 2.97 & 24.9 & 23.6 \\
\midrule
\NEW{250 Patches} & \NEW{48.4} & \NEW{0.14} & \NEW{1.74} & \NEW{1.71} & \NEW{26.6} & \NEW{21.7} \\
\NEW{500 Patches} & \NEW{28.9} & \NEW{0.21} & \NEW{1.7} & \NEW{1.45} & \NEW{26.6} & \NEW{21.1} \\
\NEW{1k Patches} & \NEW{23.5} & \NEW{0.15} & \NEW{1.26} & \NEW{1.57} & \NEW{26.3} & \NEW{22.8} \\
3k Patches & 14.4 & 0.23 & 5.1 & 8.46 & 25.0 & 23.5 \\
12k Patches & 10.6 & 0.17 & 2.1 & 3.00 & 25.7 & 23.8 \\
\midrule
\NEW{Noisy Scales} & \NEW{14.4} & \NEW{0.21} & \NEW{1.6} & \NEW{1.87} & \NEW{25.5} & \NEW{23.2} \\
\midrule   
\textbf{Ours} & 9.00 & 0.21 & 1.5 & 2.08 & 25.7 & 23.9 \\
\bottomrule
\end{tabular}
\end{table}

\section{conclusion}

We have presented a novel approach for learning a multiscale image representation from a collection of low-resolution, unstructured images.
Our method enhances an alias-free generator with progressive Fourier features distributed across various layers.
Furthermore, we have developed techniques to stabilize training and guarantee scale consistency.
The strength of our method is demonstrated in both multi-scale generative modeling and pseudo-reconstruction of scale spaces from unstructured patches.
For the first time, our neural representation achieves zoom-in factors of up to 256x, opening up a new way for efficient modeling of multi-scale images.

\begin{acks}
The authors thank Bartosz Wojczy{\'n}ski for providing the Milkyway data, as well as Joachim Weickert and Pascal Peter for early discussions.
This research was partially funded by the ERC Advanced Grant FUNGRAPH (https://fungraph.inria.fr), No 788065 and an academic gift from Meta.
\end{acks}

\bibliographystyle{ACM-Reference-Format}
\bibliography{paper}

\end{document}


\title{Learning Images Across Scales Using Adversarial Training -- Supplemental Document}

\author{Krzysztof Wolski}
\orcid{0000-0003-2290-0299}
\affiliation{%
 \institution{Max-Planck-Institut für Informatik}
 \city{Saarbrücken}
 \country{Germany}}
\email{kwolski@mpi-inf.mpg.de}

\author{Adarsh Djeacoumar}
\orcid{0009-0008-1919-450X}
\affiliation{%
 \institution{Max-Planck-Institut für Informatik}
 \city{Saarbrücken}
 \country{Germany}}
\email{adjeacou@mpi-inf.mpg.de}

\author{Alireza Javanmardi}
\orcid{0009-0008-4926-1566}
\affiliation{%
 \institution{Max-Planck-Institut für Informatik}
 \city{Saarbrücken}
 \country{Germany}}
\email{alireza.javanmardi@dfki.de}

\author{Hans-Peter Seidel}
\orcid{0000-0002-1343-8613}
\affiliation{%
 \institution{Max-Planck-Institut für Informatik}
 \city{Saarbrücken}
 \country{Germany}}
\email{hpseidel@mpi-sb.mpg.de}

\author{Christian Theobalt}
\orcid{0000-0001-6104-6625}
\affiliation{%
 \institution{Max-Planck-Institut für Informatik}
 \city{Saarbrücken}
 \country{Germany}}
\email{theobalt@mpi-inf.mpg.de}

\author{Guillaume Cordonnier}
\orcid{0000-0003-0124-0180}
\affiliation{%
 \institution{Inria, Université Côte d'Azur}
 \city{Sophia-Antipolis}
 \country{France}}
\email{guillaume.cordonnier@inria.fr}

\author{Karol Myszkowski}
\orcid{0000-0002-8505-4141}
\affiliation{%
 \institution{Max-Planck-Institut für Informatik}
 \city{Saarbrücken}
 \country{Germany}}
\email{karol@mpi-inf.mpg.de}

\author{George Drettakis}
\orcid{0000-0002-9254-4819}
\affiliation{%
 \institution{Inria, Université Côte d'Azur}
 \city{Sophia-Antipolis}
 \country{France}}
\email{george.drettakis@inria.fr}

\author{Xingang Pan}
\orcid{000-0002-5825-9467}
\affiliation{%
 \institution{Nanyang Technological University}
 \city{Singapore}
 \country{Singapore}}
\email{xingang.pan@ntu.edu.sg}

\author{Thomas Leimkühler}
\orcid{0009-0006-7784-7957}
\affiliation{%
 \institution{Max-Planck-Institut für Informatik}
 \city{Saarbrücken}
 \country{Germany}}
\email{thomas.leimkuehler@mpi-inf.mpg.de}

\maketitle

\section{Stitching with Photoshop}

\begin{figure*}
    \centering
    \includegraphics[width=\linewidth]{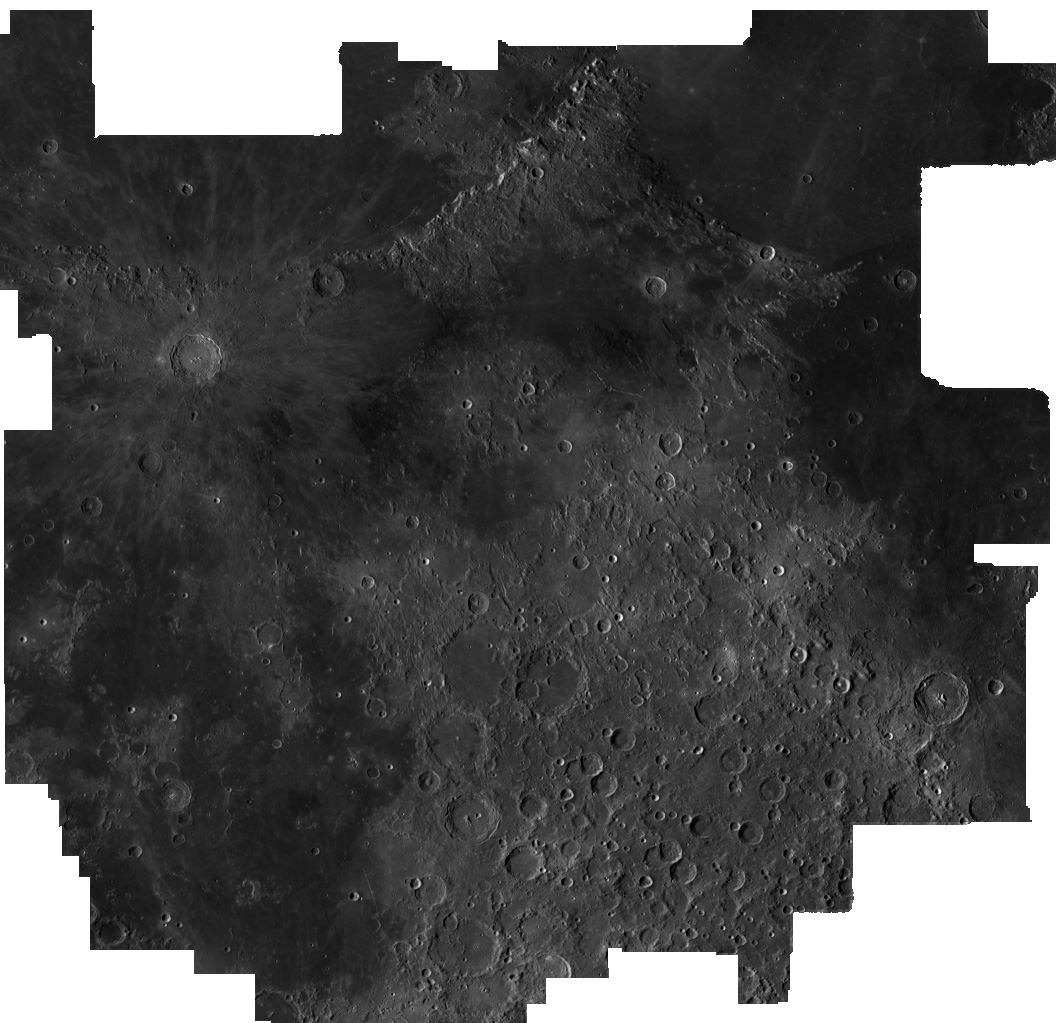}
    \caption{Image stitching with Photoshop.}
    \label{fig:photoshop}
\end{figure*}

In \refFig{photoshop}, we show the best result we could obtain using automatic image stitching with Adobe Photoshop. 
For this to work, we had to restrict the data to 420 patches at $\scale=2$, which constitutes only a tiny fraction of our data and is four scales coarser than what our approach can deliver.

\section{Baselines for Generation}

\subsection{AnyresGAN \cite{chai2022any}}
As this method, different from our setting, requires input images at different resolutions, we modify our dataset extraction procedure.
Specifically, we synthesize significantly larger images patches, from which the original implementation can extract the patches needed for training.
Notice how this provides the method with more context than our method requires.

\subsection{PULSE \cite{menon2020pulse}}
We first train a vanilla StyleGAN3~\cite{karras2021alias} model on all data patches.
To synthesize a zooming-in sequence, we first generate a random sample from this model.
\NEW{We then perform GAN inversion to create a new sample that corresponds to a 2x zoomed-in image using the PULSE~\cite{menon2020pulse} framework on the central crop of the original sample.
We repeat this process, until the required scale range is obtained.
Continuous zooming is achieved using scaling and linear blending of the obtained sample sequence.
}

\section{Dataset Statistics}

\NEW{In Tab.~\ref{tab:dataset_statistics} we report detailed statistics on our datasets. We list the number of patches per scale interval, as well as the resulting density of patches. Density is reported as the expected number of patches that overlap a single pixel from an image that has the highest possible resolution in the scale interval.}

\begin{table*}[]
    \centering
    \caption{\NEW{Dataset details. For each scale interval, we list the number of patches (\#Patch.) and the number of images per pixel (Dens.).}}
    \renewcommand{\tabcolsep}{0.115cm}
    \label{tab:dataset_statistics}
    
    \begin{tabular}{lrrrrrrrrrrrrrrrr} 
        \multirow{2}{*}[-0.5ex]{Dataset}  & \multicolumn{2}{c}{Scale 0} &
        \multicolumn{2}{c}{Scale 1} &
        \multicolumn{2}{c}{Scale 2} &
        \multicolumn{2}{c}{Scale 3} &
        \multicolumn{2}{c}{Scale 4} &
        \multicolumn{2}{c}{Scale 5} &
        \multicolumn{2}{c}{Scale 6} &
        \multicolumn{2}{c}{Scale 7} \\
        \cmidrule(lr){2-17}
        & \footnotesize{\#Patch.} & \footnotesize{Dens.} &
        \footnotesize{\#Patch.} & \footnotesize{Dens.} &
        \footnotesize{\#Patch.} & \footnotesize{Dens.} &
        \footnotesize{\#Patch.} & \footnotesize{Dens.} &
        \footnotesize{\#Patch.} & \footnotesize{Dens.} &
        \footnotesize{\#Patch.} & \footnotesize{Dens.} &
        \footnotesize{\#Patch.} & \footnotesize{Dens.} &
        \footnotesize{\#Patch.} & \footnotesize{Dens.}
        \\
        \midrule

\dataset{Milkyway} & 0.5k & 276.23 & 1.5k & 204.62 & 4.k & 136.38 & 30k & 254.30 & 30k &  63.36 & 30k & 15.86 & - & - & - & -   \\     
\dataset{Moon} & 0.5k & 276.23 & 1.5k & 204.62 & 4.k & 136.38 & 30k & 254.30 & 30k &  63.36 & 30k & 15.86 & - & - & - & -   \\
\dataset{Himalayas} & 0.5k & 276.23 & 1.5k & 204.62 & 4.k & 136.38 & 30k & 254.30 & 30k &  63.36 & 30k & 15.86 & 30k & 3.96 & 30k & 0.99   \\
\dataset{Spain} & 0.5k & 276.23 & 1.5k & 204.62 & 4.k & 136.38 & 30k & 254.30 & 30k &  63.36 & 30k & 15.86 & 30k & 3.96 & 30k & 0.99   \\
\dataset{Rembrandt} & 0.5k & 276.23 & 1.5k & 204.62 & 4.k & 136.38 & 30k & 254.30 & 30k &  63.36 & 30k & 15.86 & 30k & 3.96 & 30k & 0.99   \\

\midrule

\dataset{MilkywayGen} & 30k & - & 30k & - & 30k & - & 30k & - & - & - & - & - & - & - & - & -   \\  
\dataset{MoonGen} & 30k & - & 30k & - & 30k & - & 30k & - & - & - & - & - & - & - & - & -   \\  
\dataset{HimalayasGen} & 30k & - & 30k & - & 30k & - & 30k & - & - & - & - & - & - & - & - & -   \\
\dataset{SpainGen} & 30k & - & 30k & - & 30k & - & 30k & - & - & - & - & - & - & - & - & -   \\
\dataset{RembrandtGen} & 30k & - & 30k & - & 30k & - & 30k & - & - & - & - & - & - & - & - & -   \\
\dataset{Sunflowers} & 30k & - & 60k & - & 60k & - & 35k & - & - & - & - & - & - & - & - & -   \\
\dataset{Bricks} & 25k & - & 67k & - & 82k & - & 60k & - & - & - & - & - & - & - & - & -   \\

\midrule

\dataset{Milkyway 12k} & 62 & 33.59 & 187 & 25.28 & 500 & 16.91 & 3750 & 31.70 & 3750 & 7.92 & 3750 & 1.98 & - & - & - & -   \\
\dataset{Milkyway 3k} & 15 & 8.14 & 46 & 6.22 & 125 & 4.23 & 937 & 7.92 & 937 & 1.98 & 937 & 0.50 & - & - & - & -   \\
\dataset{Milkyway 1k} & 5 & 2.70 & 15 & 2.02 & 41 & 1.39 & 312 & 2.64 & 312 & 0.66 & 312 & 0.16 & - & - & - & -   \\
\dataset{Milkyway 0.5k} & 2 & 1.08 & 7 & 0.95 & 20 & 0.68 & 156 & 1.32 & 156 & 0.33 & 156 & 0.08 & - & - & - & -   \\
\dataset{Milkyway 0.25k} & 3 & 1.62 & 3 & 0.41 & 10 & 0.34 & 78 & 0.66 & 78 & 0.16 & 78 & 0.04 & - & - & - & -   \\    
        \bottomrule
    \end{tabular}
\end{table*}

\bibliographystyle{ACM-Reference-Format}
\bibliography{paper}